# Tuning the metal-insulator transition in epitaxial SrVO$_3$ films by uniaxial strain

Changan Wang[*,1,2,3] Hongbin Zhang[*,4] Kumar Deepak,[5] Chao Chen,[6] Arnaud Fouchet,[5] Juanmei Duan,[1,2,3] Donovan Hilliard,[1] Ulrich Kentsch,[1] Deyang Chen,[6,7] Min Zeng,[6] Xingsen Gao,[6] Yu-Jia Zeng,[3] Manfred Helm,[1,2] Wilfrid Prellier,[5] and Shengqiang Zhou[1]

1) Helmholtz-Zentrum Dresden-Rossendorf, Institute of Ion Beam Physics and Materials Research, Bautzner Landstr. 400, 01328 Dresden, Germany
2) Technische Universität Dresden, D-01062 Dresden, Germany
3) Shenzhen Key Laboratory of Laser Engineering, College of Physics and Optoelectronic Engineering, Shenzhen University, 518060 Shenzhen, China
4) Department of Materials and Geosciences, TU Darmstadt, Darmstadt, 64287, Germany
5) Laboratoire CRISMAT, CNRS UMR 6508, ENSICAEN, Normandie Université, 6Bd Maréchal Juin, F-14050 Caen Cedex 4, France
6) Institute for Advanced Materials and Guangdong Provincial Key Laboratory of Quantum Engineering and Quantum Materials, South China Academy of Advanced Optoelectronic, South China Normal University，Guangzhou 510006, China
7) National Center for International Research on Green Optoelectronics, South China Academy of Advanced Optoelectronics, South China Normal University, Guangzhou 510006, China

**ABSTRACT**

[*] Author to whom correspondence should be addressed. e-mails: changan.wang@hzdr.de and hzhang@tmm.tu-darmstadt.de

Understanding of the metal-insulator transition (MIT) in correlated transition-metal oxides is a fascinating topic in condensed matter physics and a precise control of such transitions plays a key role in developing novel electronic devices. Here we report an effective tuning of the MIT in epitaxial SrVO$_3$ (SVO) films by expanding the out-of-plane lattice constant without changing in-plane lattice parameters, through helium ion irradiation. Upon increase of the ion fluence, we observe a MIT with a crossover from metallic to insulating state in SVO films. A combination of transport and magnetoresistance measurements in SVO at low temperatures reveals that the observed MIT is mainly ascribed to electron-electron interactions rather than disorder-induced localization. Moreover, these results are well supported by the combination of density functional theory and dynamical mean field theory (DFT+DMFT) calculations, further confirming the decrease of the bandwidth and the enhanced electron-electron interactions resulting from the expansion of out-of-plane lattice constant. These findings provide new insights into the understanding of MIT in correlated oxides and perspectives for the design of unexpected functional devices based on strongly correlated electrons.

# 1. INTRODUCTION

Correlated transition-metal oxides host a vast landscape of novel phenomena in condensed matter physics, such as multiferroicity, metal-insulator transitions (MIT), quantum magnetism and colossal magnetoresistance [1-5]. Among these materials, perovskite $SrVO_3$ (SVO) with a $3d^1$ electronic configuration is one of such typical strongly correlated systems used to study the MIT. Although SVO bulk and films have been extensively studied in the past, exhibiting a robust metallic character and an enhanced Pauli paramagnetic behavior, theoretical studies have predicted that a MIT in SVO thin films can appear as the SVO thickness is reduced to a few unit cells [6,7]. Fortunately, experimental results in ultrathin films seem to verify this prediction: a dimensional-crossover-induced MIT was observed in single SVO ultrathin films [8-11]. Similar results have also been obtained in other materials, including $LaNiO_3$, $La_{2/3}Ca_{1/3}MnO_3$ and $CaVO_3$ [12-14]. Based on the Mott-Hubbard theory, MIT can be controlled by the competition between the on-site Coulomb repulsion $U$ and the electronic bandwidth $W$ [1]. The aforementioned MIT in these films is attributed to a reduced effective $W$ with decreasing film thickness, which facilitates a Mott insulating state and forms a pseudogap at the Fermi level [8]. On the other hand, chemical substitution can also be used to study the MIT. It has been reported that a MIT is induced in SVO alloyed with La or Ca ($La_{1-x}Sr_xVO_3$ or $Ca_{1-x}Sr_xVO_3$), in which $W$ is influenced by the resultant changes in the bond angle and band length of oxygen ions with transition metal ions. Thus, the electron correlation $U/W$ is enhanced in $La_{1-x}Sr_xVO_3$ or $Ca_{1-x}Sr_xVO_3$ [15,16]. For the ultrathin films, however, one should notice that the dimensional-crossover-induced MIT experiences a thickness-dependent relaxation of strain due to the lattice mismatch with substrates, which is inevitable and is not considered in these studies, although the strain relaxation may have a great effect on the MIT of these ultrathin films. In addition, chemical substitution always induces chemical disorders in strongly correlated systems that may also influence the MIT [17]. Therefore, in-depth understanding of the origin of MIT in correlated transition-metal oxides remains challenging and is essential from both fundamental and practical points of views [18].

Heteroepitaxial strain engineering has emerged as a powerful tool to study strongly correlated systems and to explore new functionalities [19]. Effective control over lattice strain in the films has revealed many important properties, including ferromagnetism, superconductivity and ferroelectricity [20-22]. Here, we use helium (He) ion irradiation to induce "strain doping" in SVO films. Strain doping by irradiation of noble He atoms results in an uniaxial lattice expansion along the out-of-plane direction without changing the in-plane lattice constants [23-25]. For instance, Gao *et al*. reported that the out-of-plane lattice parameter in epitaxial La$_{0.7}$Sr$_{0.3}$MnO$_3$ thin films is continuously and independently manipulated by using He ion irradiation [23]. On the other hand, this approach bypasses the Poisson effect that always changes the lattice constant in all three directions and can clarify the related effects between order parameters and single degrees of freedom [23,25]. Thus, this approach may provide more underlying information on MIT in correlated transition-metal oxides.

In this work, we have applied He ion irradiation to effectively modify the out-of-plane lattice parameter of the SVO films that were epitaxially grown on the STO substrates. Transport measurements revealed the emergence of the MIT with a crossover from metallic to insulating state in SVO films as the irradiation fluence is increased. Further studies of transport and magnetoresistance of the SVO films at low temperatures verified that the observed MIT is mainly attributed to electron-electron interactions rather than disorder-induced weak localization. Moreover, the combination of density functional theory and dynamical mean field theory (DFT+DMFT) calculations also indicated the enhanced electron-electron interaction due to the decrease of bandwidth $W$ as the out-of-plane lattice is gradually expanded. These results therefore show the prospect of strain engineering by using He ion irradiation, providing a platform for modifying and exploring correlated behavior in complex oxides.

## 2. EXPERIMENTAL SECTION

SVO thin films (thickness: 50nm) were epitaxially grown on SrTiO$_3$ (001) single crystal substrate ($a_{sub}$ = 3.905 Å) by pulsed laser deposition (PLD). The deposition temperature was 650 ºC, the oxygen pressure was $1\times10^{-6}$ mbar, the KrF laser energy was 200 mJ, and the rate of 2 Hz was applied. After the deposition, these SVO films were then irradiated by He ions using an energy of 5 keV with different fluence values of $1\times10^{15}$, $1.75\times10^{15}$, $2.5\times10^{15}$, $3\times10^{15}$, and $3.5\times10^{15}$ He/cm$^2$. The sample area was scanned over by the ion beam to ensure a lateral uniform irradiation. For convenient description in article, the samples were referred to as $1\times10^{15}$, $1.75\times10^{15}$, $2.5\times10^{15}$, $3\times10^{15}$, and $3.5\times10^{15}$, respectively. Besides, the Stopping and Range of Ions in Matter (SRIM) simulations confirmed that He-ions give rise to a relatively homogeneous lattice displacement into SVO thin films [26]. The crystal structures were characterized by X-ray diffraction (XRD) (PANalytical X'Pert PRO diffractometer) using Cu K$\alpha$ radiation. The electrical transport measurements were carried out using the van der Pauw configuration in a Lakeshore Hall measurement system.

## 3. RESULTS AND DISCUSSION

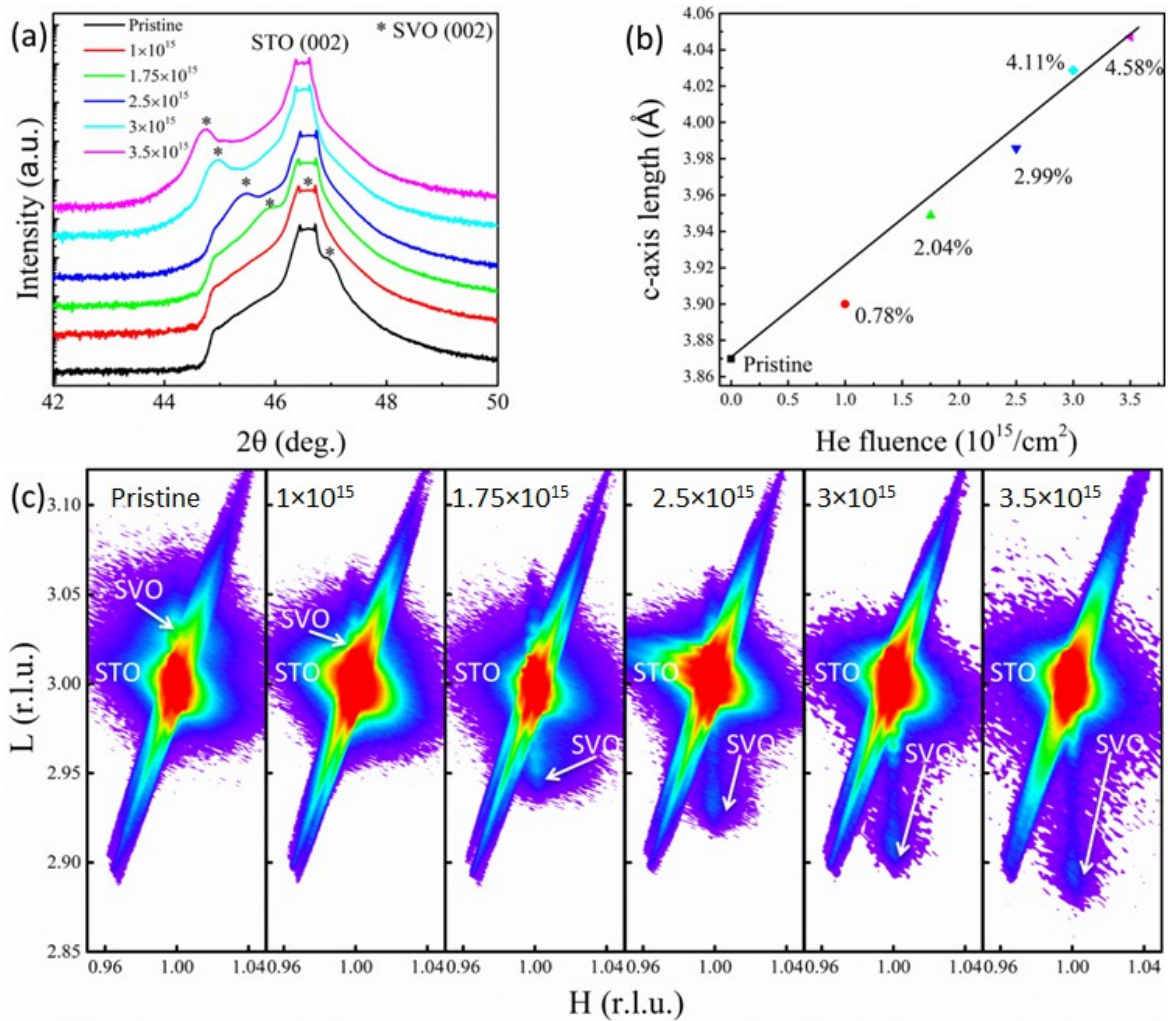

**Figure 1.** (a) XRD $\theta$-$2\theta$ scans around the (002) reflection of SVO thin films on STO substrates under different helium fluences. (b) The out-of-plane lattice constant changes as a function of the helium fluence where percentages represent the change in the c-axis relative to the pristine state. (c) Reciprocal space maps around (103) reflections for different helium fluences.

Figure 1(a) shows the XRD scans of the (002) reflection for SVO films without and with irradiation. The pristine SVO film is epitaxial to the STO substrate ($a_{sub}$ = 3.905 Å) and undergoes tensile strain from the substrate, which is in agreement with previous reports [9,10]. With the fluence of He ions increases, the position of the SVO (002) peak exhibits a systematic shift towards the low angle direction, indicating an expansion in the out-of-plane lattice parameter $c$. The $c$-axis values can be estimated from the XRD diffraction peaks and are summarized in Figure 1(b). It can be seen that the out-of-plane lattice of the SVO films increases in a continuous manner with increasing the fluence. The increase percentages with respect to the pristine one are 0.78%, 2.04%, 2.99%, 4.11%, and 4.58%, corresponding to $1\times10^{15}$, $1.75\times10^{15}$, $2.5\times10^{15}$, $3\times10^{15}$, and $3.5\times10^{15}$, respectively.

Figure 1(c) presents the reciprocal space mappings (RSMs) for all samples around (103) asymmetric reflections. One clearly sees that the in-plane lattice parameter of the pristine SVO film is the same as that of the STO substrate, further demonstrating that the SVO films are epitaxially grown on the substrate. When the fluence increases, H values of the SVO diffraction peaks are found to remain the same as that of the STO substrate due to the constraint from the substrate itself, while L shifts towards lower values due to the out-of-plane lattice expansion. Even after the irradiation with the highest fluence of $3.5\times10^{15}$ He/cm$^2$, the in-plane lattice is still epitaxially locked to the substrate. The strain induced by the irradiation is relieved along the out-of-plane direction and therefore only expands the *c*-axis parameter in SVO films [23]. Different from other methods that often involve changes in all three lattice directions, these results demonstrate that a single out-of-plane lattice parameter is independently controlled by the strain arising from He ion irradiation.

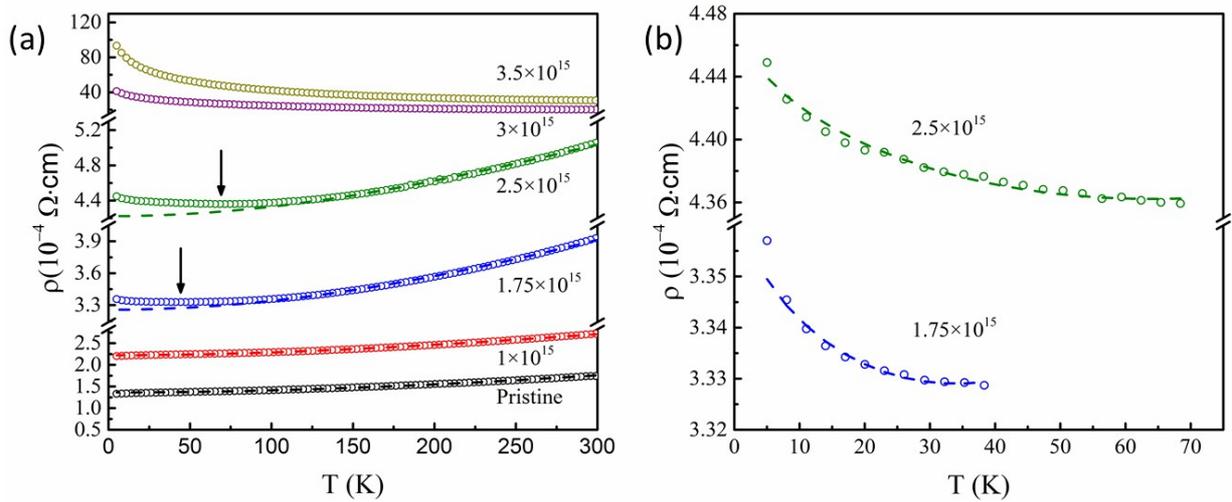

**Figure 2.** (a) Electrical resistivity vs. temperature for SVO films with increasing fluence. Arrows indicate the minima of resistivity (or MITs). The dashed lines are the fitting of data in the metallic region using Equation 1 and Equation 2. (b) Increase of resistivity on the insulating side for SVO films irradiated with $1.75\times10^{15}$ and $2.5\times10^{15}$ He/cm$^2$. The dashed lines are the fitting of data in the insulating region according to Equation 3.

In order to explore transport properties of the SVO films, Figure 2(a) shows the evolution of resistivity versus temperature with increasing fluence of He ions. The pristine film exhibits metallic behavior (dρ/dT > 0) in the investigated temperature range, being consistent with previous reports [10,14]. With increasing the fluence of irradiation, the resistivity increases compared with the pristine sample, which is attributed to the increased out-of-plane lattice and disorder induced by irradiation. In particular, the films with intermediate fluence display a MIT characterized by an upturn in resistivity. The characteristic MIT temperatures ($T_{MI}$) are

~40 K and ~68 K for fluences $1.75\times10^{15}$ and $2.5\times10^{15}$ He/cm$^2$, respectively, and are indicated by arrows in Figure 2(a). $T_{MI}$ shifts to a higher temperature upon increasing the fluence. Finally, insulating behavior (dρ/dT < 0) is found at all temperatures for the films with the fluence larger than $3\times10^{15}$ He/cm$^2$. All these results therefore obviously indicate that the SVO films undergo a MIT with a crossover from a metallic to insulating state, accompanying the expansion of the out-of-plane lattice parameter due to the irradiation.

To further explore the effect of irradiation on electrical transport behavior, we first describe the metallic state. Vanadate SVO is considered to be a Fermi liquid (FL) [9], for which the temperature dependence (T) of resistivity follows the equation with an exponent n = 2

$$\rho_{ideal}(T)=\rho_0+AT^2 \quad (1)$$

where the residual resistivity $\rho_0$ is due to the electron-impurity scattering caused by defects, and A is a measure of the strength of electron-electron interactions. However, Equation 1 does not fully describe experimental data [27]. ρ(T) can only be well described when we introduce the resistivity saturation ($\rho_{SAT}$) as follows [28,29]

$$\rho^{-1}(T)=\rho_{ideal}^{-1}+\rho_{SAT}^{-1} \quad (2)$$

where ($\rho_{SAT}$) is a parallel resistor, that is applied to a wide range of materials exhibiting the resistance saturation [28,30]. Here, the dashed lines in Figure 2(a) are the fitting results by Equation 1 and 2, which can well describe all SVO films in the metallic state. The fitting parameters of the SVO films are listed in Table 1. $\rho_0$ continuously increases, which is attributed to the increased defects induced by the irradiation, while $\rho_{SAT}$ is essentially fluence-independent, which is often related to the maximum resistivity [28,31]. More remarkable, when the fluence is increased, the coefficient A that quantifies the electron-electron interaction is enhanced by approximately 8 times, from $5.53\times10^{-10}$ Ω·cm/K$^2$ for the pristine to $4.29\times10^{-9}$ Ω·cm/K$^2$ for sample $2.5\times10^{15}$, which is believed to account for the appearance of the MIT.

Table 1. Transport properties of SVO films with different irradiation fluence

| Sample | $T_{MI}$ (K) | $\rho_0$ (Ω·cm) | A (Ω·cm/K$^2$) | $\rho_{SAT}$ (Ω·cm) |
|---|---|---|---|---|
| Pristine | – | $1.80\times10^{-4}$ | $5.53\times10^{-10}$ | $5.56\times10^{-4}$ |
| $1\times10^{15}$ | – | $4.10\times10^{-4}$ | $2.25\times10^{-9}$ | $4.86\times10^{-4}$ |
| $1.75\times10^{15}$ | 40 | $5.62\times10^{-4}$ | $2.54\times10^{-9}$ | $7.75\times10^{-4}$ |
| $2.5\times10^{15}$ | 68 | $8.29\times10^{-4}$ | $4.29\times10^{-9}$ | $5.60\times10^{-4}$ |
| $3\times10^{15}$ | >300 | - | - | - |
| $3.5\times10^{15}$ | >300 | - | - | - |

Figure 2(b) presents the resistivity at the insulating side of the MIT for $1.75\times10^{15}$ and $2.5\times10^{15}$ films, which shows the resistivity upturns below $T_{MI}$ in both samples. This phenomenon is attempted to be interpreted by quantum corrections of the conductivity (QCC) [32]. In this three-dimensional system (3D), the temperature (T) dependence of the resistivity (ρ) can be described by [10,32]:

$$\rho(T)=\frac{1}{\sigma_0+a_1 T^{P/2}+a_2 T^{1/2}}+b T^2 \quad (3)$$

where $a_1 T^{P/2}$ is the 3D weak localization (WL) contribution ($P$ = 2 or 3 is the electron-electron interactions or electron-phone scattering, respectively) [33], while $a_2 T^{1/2}$ stands for the renormalized electron-electron interaction (REEI), representing the density of state correction at the Fermi energy. Another term $bT^2$ accounts for the classic low temperature dependence of the resistivity. Clearly, the resistivity at the insulating side of the MIT in both samples can be well fitted with Equation 3 (see dashed lines in Figure 2(b)), which is consistent with previous observations in SVO films [10]. Although QCC successfully describes the insulating region below $T_{MIT}$, it is not clear whether WL or REEI is the main contribution to the resistivity upturn (or QCC) [33]. Since both effects have similar temperature-dependent behaviors in the 3D case, $\Delta\sigma \sim T^{1/2}$ or $\Delta\sigma \sim T^{P/2}$ with $P$=2-3 for REEI or WL, respectively. To clarify the driving force for this phenomenon in our SVO systems, the field dependence of the resistivity at low temperature needs to be considered [33].

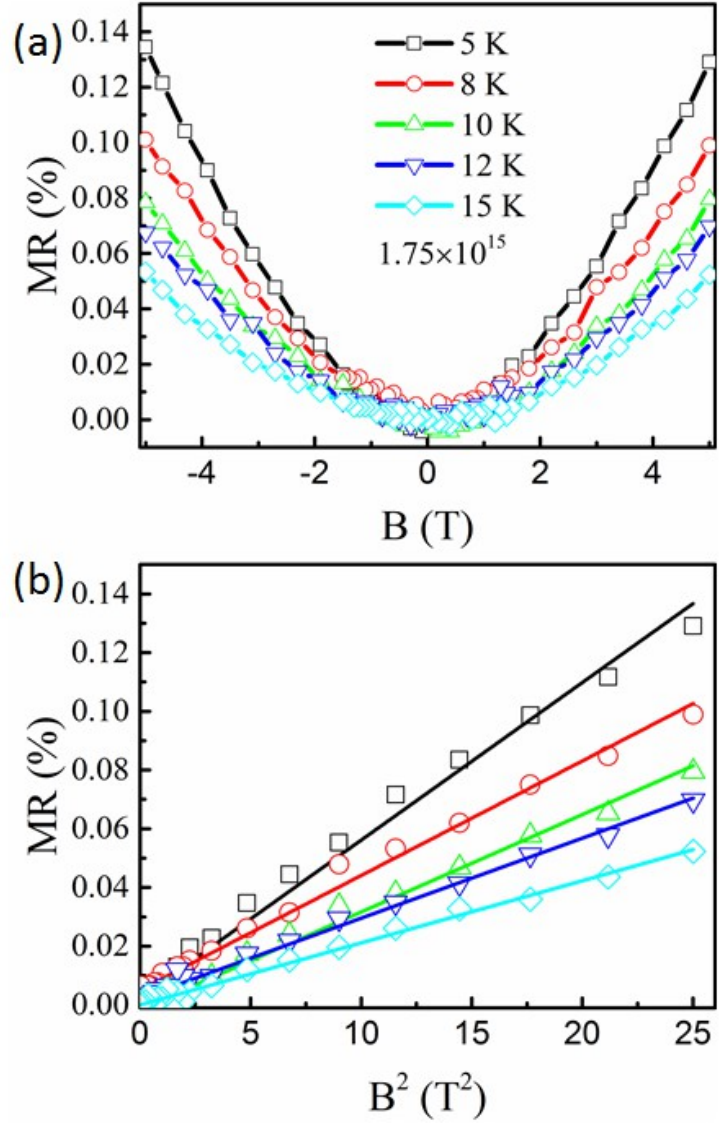

**Figure 3.** (a) The field dependence of the magnetoresistance MR=[R(*B*)-R(0)]/R(0) in the SVO sample irradiated with fluence of $1.75\times10^{15}$ He/cm$^2$ at different temperature. (b) The MR is plotted versus the square of the magnetic field. The linear fits to the data are also shown.

Figure 3(a) shows magnetoresistance (MR, defined as [R(B)-R(0)/R(0)]) at different temperatures for the SVO sample irradiated with the fluence of $1.75\times10^{15}$ He/cm$^2$. A positive MR is observed which decreases with increasing temperature. Moreover, the MR is found to be roughly proportional to B$^2$ for each temperature, and linear fits are also shown in Figure 3(b). These observations rule out that the WL is the main contribution to the insulating behavior at low temperatures [9]. This is because the effect of the WL leads to a negative MR originating from the suppression of self-interference effects by applying a magnetic field [9,33]. Therefore, the observed MIT in our systems is mainly attributed to the electron-electron interactions rather than the WL.

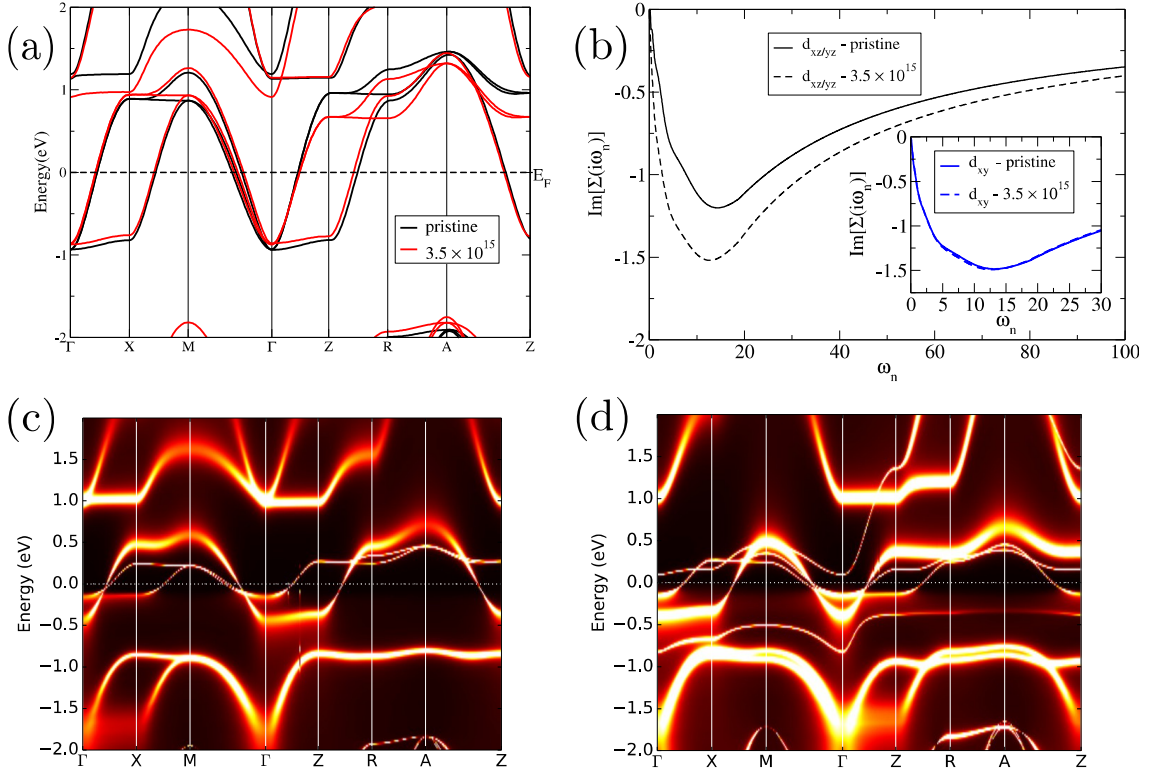

**Figure 4.** (a): DFT band structure (a) of the pristine (black, c = 3.87 Å) and 3.5×10$^{15}$ (red, c = 4.05 Å) samples. (b): Imaginary part of the DFT+DMFT self energies for the degenerate $d_{xz}$ and $d_{yz}$ orbitals for the pristine (solid line) and 3.5×10$^{15}$ (dashed line) samples. The inset shows the self energy for the $d_{xy}$ orbitals. (c) and (d) show the DFT+DMFT spectral functions for the pristine and 3.5×10$^{15}$ samples, respectively. The horizontal dashed lines in (a), (c) and (d) indicate the Fermi energy.

To shed light on the physical origin of the MIT in SVO, we performed DFT+DMFT calculations. Our calculations were carried out using the projection-embedding implementation based on the WIEN2k package [34,35], where the auxiliary impurity problem is solved using the continuous-time quantum Monte Carlo method [36]. The local density approximation is applied for the exchange correlation potential as parameterized by Perdew-Wang [37]. All the d-orbitals of V atoms are considered as correlated orbitals, while the hybridization and self-energies are computed in an energy window of ±10 eV around the Fermi energy, resulting in U = 8.0 eV and J = 0.8 eV. The DFT+DMFT calculations are all done at 50 K, where the in-plane lattice constants are fixed to 3.905 Å corresponding to that of SrTiO$_3$ substrates, with the out-of-plane lattice constant following the experimental XRD measurements.

Following Figure 4(a), the bandwidth W of the $d_{xz}$ and $d_{yz}$ bands is reduced upon irradiation, *i.e.*, the band width W is inversely proportional to the lattice constant along the c-axis for the $d_{xz}$ and $d_{yz}$ orbitals, while that of the $d_{xy}$ bands remains almost a constant due to the fixed in-plane lattice constants. The reduced bandwidth leads to enhanced density of states (DOS) at the Fermi energy (not shown), leading to a possibly enhanced conductivity following the Drude model which would be in contradiction to the experimental observation. We found that the origin of the reduced conductivity upon irradiation can be attributed to enhanced scattering due to electronic correlations. Considering the renormalization factor

$$Z = \left(1 - \frac{dIm\,\Sigma}{d\omega}\bigg|_{\omega=0}\right)^{-1}$$

, where $\Im\Sigma$ denotes the DMFT self-energy at Matsubara frequency $\omega_n$ (as shown in Figure 4(b)), it is reduced from 0.77 (pristine) to 0.56 ($3.5\times10^{15}$) for both $d_{xy}$ and $d_{yz}$ orbitals, whereas it remains a constant of 0.57 for the $d_{xy}$ orbital. Thus, the expansion of the c-axis lattice constant upon irradiation causes orbital-selective enhancement of the electronic correlations, resulting in the abnormal temperature dependence of the resistivity. However, there is no orbital-selective Mott transition [38]. For instance, for the irradiated samples there exists an abnormal MIT-like dependence of resistivity with respect to the temperature (Figure 2(a)), an interesting question is whether a finite band gap is opened. According to the spectral functions for the pristine (Figure 4(c)) and $3.5\times10^{15}$ (Figure 4(d)) samples obtained in our DFT+DMFT calculations, there is no gap opening, whereas the bandwidth for the $d_{xy}$ and $d_{yz}$ bands is renormalized by a factor of 0.25 compared to the DFT electronic structure (cf. Figure 4(a)). It is noted that for the $3.5\times10^{15}$ sample, the hybridization of the $t_{2g}$ bands with the other bands has been significantly changed (Figure 4(d)). Thus, the abnormal temperature dependence of the resistivity in SVO films can be mainly attributed to electronic correlations, whose strength is tunable via He ion irradiation which is a matured technique. This study offers an effective method to explore and control MIT in correlated materials, allowing for the design of future electronic applications.

## 4. CONCLUSIONS

We have successfully manipulated the out-of-plane lattice parameter in SVO films without altering other degrees of freedom by using He ion irradiation. The transport study indicates a metal-insulator transition (MIT) with a crossover from the metallic to insulating state in SVO films as the fluence is increased. The observed MIT can be attributed to the decrease of the electronic bandwidth $W$ due to the expansion of the out-of-plane lattice parameter induced by the irradiation, and then enhanced electron-electron interactions. The detailed investigations of transport, magnetoresistance and DFT+DMFT calculations in the insulating regime also reveal an obvious signature of strong electron-electron interactions, which is the dominant force in driving the observed MIT in SVO films. These results clearly pave a way to tune correlated behavior and develop novel functional devices based on transition-metal oxides.

**ACKNOWLEDGMENTS**

Support by the Ion Beam Center at Helmholtz-Zentrum Dresden-Rossendorf is gratefully acknowledged. C. A. Wang thanks China Scholarship Council (File No. 201606750007) for financial supports. S. Zhou acknowledges financial support from the German Research Foundation (ZH 225/10-1). This work was also supported by the Shenzhen Science and Technology Project under Grant Nos. JCYJ20170412105400428 and KQJSCX20170727101208249. D. C. thanks the financial support from the National Natural Science Foundation of China (Grant Nos. 11704130 and U1832104).